\documentclass[amsmath,amssymb,floatfix,prb]{revtex4}
\usepackage{graphicx}
\usepackage{color}
\usepackage{amssymb,amsmath}

\begin{document}

\title{\bf Reply to 
Comment on ``Critical analysis of a variational method used to describe molecular 
electron transport''}

\author{Ioan B\^aldea}
\email{ioan.baldea@pci.uni-heidelberg.de}
\altaffiliation[Also at ]{National Institute for Lasers, Plasma, 
and Radiation Physics, ISS, POB MG-23, RO 077125 Bucharest, Romania.}
\author{Horst K\"oppel} 
\affiliation{Theoretische Chemie,
Physikalisch-Chemisches Institut, Universit\"{a}t Heidelberg, Im
Neuenheimer Feld 229, D-69120 Heidelberg, Germany}
\begin{abstract}
We show that the failure of the Delaney-Greer (DG) variational ansatz 
for transport demonstrated
by us in Phys.\ Rev.\ B {\bf 80}, 165301 (2009) (I)
is not related to an unsuitable constraint that 
prevents a broken time-reversal symmetry or to real orbitals, as DG  
incorrectly claim.
The complex orbitals suggested by them as a way-out solution 
merely represent a particular case of the general case considered by us in I,
which do not in the least affect our conclusion. 
In conjunction with the
issues raised by the DG's Comment, we show 
that the DG Wigner conditions can erroneously constrain
\emph{outgoing} and not incoming charge carriers 
and present an example revealing that the
sign of the ``momentum'' $P$ of the Wigner function $f(q, P)$
is not necessarily associated with the direction of motion in the real world.
We also discuss a general reason why transport approaches which, like the DG's, are solely
based on information on an isolated nanocluster are incorrect.
\end{abstract}
\pacs{73.63.-b, 73.23.-b}
\keywords{molecular electronics, nanotransport, Wigner function, variational principles}
\date{\today}
\maketitle
\renewcommand{\topfraction}{1}
\renewcommand{\bottomfraction}{1}
\renewcommand{\textfraction}{0}
\section{Introduction}
\label{sec-introduction}
In their Comment, \cite{GreerComment:10} Delaney and Greer (DG) claim 
that the unphysical result (current $J=0$) presented by us in 
Ref.~\onlinecite{Baldea:2009c} (hereafter, I) as evidence against 
their variational approach \cite{DelaneyGreer:04a}
is due to the fact (i) that our constraints (electronic populations)
prevent a broken time-reversal symmetry, and (ii) that we used
real orbitals, which concomitantly constrain incoming and outgoing electrons.
Further, DG accept that our idea to constrain populations ``is an interesting alternative to 
the Wigner function'' (WF), but (iii) argue that to get $J\neq 0$ 
it is essential to use complex orbitals, like those obtained by
applying periodic boundary conditions (pBCs) 
for electrodes. 
By using the superscript $DG$ and by referring to their work,\cite{DelaneyGreer:04a}
which we amply criticized \cite{Baldea:2008b,Baldea:2009c,Baldea:2010g}
they aim to convey the false impression that they used such complex orbitals previously.

Responding briefly, claim (i) is incorrect, claim (ii) has absolutely no real basis, 
while claim (iii) does not mender their variational approach; the complex orbitals
discussed by them solely represent a particular case of the general case of I
and do not in the least change the unphysical linear response current $J=0$.
\section{Rebuttal of the criticism that population constraints prevent a broken time-reversal symmetry}
\label{sec-time-reversal}
DG argue that the unphysical result ($J=0$) 
deduced in I is the consequence of our unsuitable choice 
of constraints, namely the particle number operators. 
They (incorrectly) claim that 
constraining {\em real} particle number operators does not break 
time-reversal symmetry, and for this very reason
the current vanishes even beyond the linear response limit examined in I.

We shall immediately show that this is not true: our constraints 
\emph{allow} a broken time-reversal symmetry. The fact that this symmetry is 
\emph{not} broken is the \emph{result} of the defective DG variational ansatz (constrained 
energy minimization at zero temperature).
Below, we shall only discuss the 
particular constraint $Q_{\kappa} = n_{\kappa} \equiv \alpha_{\kappa}^{\dagger} \alpha_{\kappa}$ 
mentioned in the Comment.
Let us consider a general many-body state,
$\left\vert \Psi\right \rangle =  \sum_{n} A_n \left\vert\Psi_n\right\rangle $,
expanded in terms of the complete set of the real eigenstates 
of the Hamiltonian $H$ of the finite transport cluster 
without external bias ($V=0$), 
$H\vert\Psi_n\rangle = E_n \vert\Psi_n\rangle $. Notice that this \emph{exact}
expansion is just 
in the spirit of the DG approach, which is based on a
configuration interaction (CI), albeit approximate 
expansion [cf.~Eq.~(1) of Ref.~\onlinecite{DelaneyGreer:04a}].
The expansion coefficients $A_n$ \emph{are} allowed to be complex.
Accepting the DG challenge, we consider now the general response 
instead of the linear limit of I.
We do no more single out the ground state $\vert \Psi_0\rangle $, unlike e.~g., Eq.~(10)
of I, and consider the general matrix elements of the relevant operators 
[external bias Hamiltonian, $W$, and current operator at site $q$,
$j_{q} = i e t_q (a_{q}^{\dagger} a_{q + 1} - a_{q+1}^{\dagger} a_{q})$]
\begin{eqnarray}
\label{eq-Psi}
\mathcal{W}_{n m} & \equiv & \left\langle \Psi_n\left\vert W \right\vert \Psi_m\right\rangle = \mathcal{W}_{m n};
\mathcal{X}_{n m}^{\kappa} \equiv \left\langle \Psi_n\left\vert \alpha_{\kappa}^{\dagger} \alpha_{\kappa}\right\vert \Psi_m\right\rangle = \mathcal{X}_{m n}^{\kappa}, \\
\mathcal{J}_{n m}^{q} & \equiv & 
- i \left\langle \Psi_n\left\vert 
\left(j_q - j_{q_0}\right)
\right\vert \Psi_m\right\rangle 
\equiv \mathcal{I}_{n m}^{q} - \mathcal{I}_{n m}^{q_0}
= - \mathcal{J}_{m n}^{q} ,
\end{eqnarray}
where $q_0$ is an arbitrary fixed site. 
(Unless otherwise specified, we use throughout the notations and definitions of I 
and ignore the spin for simplicity.)
Notice that all the above calligraphic symbols denote real quantities. 
To determine the steady-state current, 
the DG variational ansatz prescribes a constrained energy minimization
\begin{equation}
\label{eq-min}
\sum_{n,m} A_n^{\ast} A_m 
\left[ \left(E_n - \omega\right) \delta_{n m}  + \mathcal{W}_{n m}
- \sum_{\kappa} \lambda_{\kappa} \mathcal{X}_{n m}^{\kappa}
- i \sum_{q\neq q_0} \chi_{q} \mathcal{J}_{n m}^{q} \right] = \min ,
\end{equation}
where $\omega$, $\lambda_{\kappa}$, and $\chi_{q}$ are real Lagrange multipliers.
The constraints are (see I):
\begin{eqnarray}
\label{eq-Q-k}
& & \sum_{n,m} A_n^{\ast} A_m \mathcal{X}_{n m}^{\kappa} = \mathcal{X}_{0 0}^{\kappa} , \\
\label{eq-j-q}
& & \sum_{n,m} A_n^{\ast} A_m \mathcal{J}_{n m}^{q} = 0 , \\
\label{eq-norm}
& & \sum_{n,m} A_n^{\ast} A_m = 1 . 
\end{eqnarray}
The minimization with respect to $A_n$ and $A_n^{\ast}$ of Eq.~(\ref{eq-min}) yields 
\begin{equation}
\label{eq-An}
\sum_{m}  A_m  \left[ \left(E_n - \omega\right) \delta_{n m}  + \mathcal{W}_{nm}
- \sum_{\kappa} \lambda_{\kappa} \mathcal{X}_{n m}^{\kappa}
- i \sum_{q\neq q_0} \chi_{q} \mathcal{J}_{n m}^{q} \right] = 0 
\end{equation}
and 
\begin{equation}
\label{eq-An-ast}
\sum_{m}  A_m^{\ast}  \left[ \left(E_n - \omega\right) \delta_{n m}  + \mathcal{W}_{nm}
- \sum_{\kappa} \lambda_{\kappa} \mathcal{X}_{n m}^{\kappa}
+ i \sum_{q\neq q_0} \chi_{q} \mathcal{J}_{n m}^{q} \right] = 0 ,
\end{equation}
respectively. Notice the reversed sign of the last term in the square 
parentheses of Eqs.~(\ref{eq-An}) and (\ref{eq-An-ast}). It is due the 
antisymmetry expressed in Eq.~(\ref{eq-j-q}), which is related to the fact that 
the matrix elements the Hermitian current 
operator are purely imaginary, i.~e.,
$\mathcal{I}_{nm}^{q}$ are real.
Rephrasing, this reflects the fact that if 
the state $\vert \Psi\rangle = \sum_n A_n \vert \Psi_n\rangle$ corresponds to a current $+J$,
$\vert \Psi^{\ast}\rangle = \sum_n A_n^{\ast} \vert \Psi_n\rangle$ 
corresponds to a current $-J$. 
To determine the expansion coefficients (and the Lagrange multipliers),
one must solve Eqs.~(\ref{eq-Q-k}) --- (\ref{eq-An-ast}), and then
the position-independent current can be computed
\begin{equation}
\label{eq-J}
J \equiv J_q = i \sum_{n,m}  A_n^{\ast}  A_m \mathcal{I}_{nm}^{q} = 
\frac{i}{2} \sum_{n,m}  \left( A_n^{\ast}  A_m - A_n A_m^{\ast} \right) \mathcal{I}_{nm}^{q}.
\end{equation}
As visible in Eqs.~(\ref{eq-An}) and (\ref{eq-An-ast}),
$A_n$ and $A_n^{\ast}$ obey \emph{different} equations.  
So, the $A_n$'s are 
\emph{allowed} to be complex, hence $\vert\Psi\rangle$ is also 
\emph{allowed} to be complex. Constraining populations, Eq.~(\ref{eq-Q-k}), 
\emph{allows} a broken time-reversal symmetry.
To see whether the $A_n$'s \emph{are} indeed complex 
(thence whether a current $J$, Eq.~(\ref{eq-J}), \emph{can} indeed flow), 
one has to solve the set of Eqs.~(\ref{eq-Q-k}) --- (\ref{eq-An-ast}).
Beyond linear response, Eqs.~(\ref{eq-Q-k}) --- (\ref{eq-An-ast}) are 
coupled nonlinear equations, and the solution is \emph{not} necessarily unique.
In I, we do \emph{solve} this problem for linear response, show 
that the solution \emph{is} unique, and the 
unphysical result $J=0$ is the \emph{outcome} of these calculations.
Throughout our critical analysis of the DG approach,
\cite{Baldea:2008b,Baldea:2009c,Baldea:2010g} to be on the safe side, 
we confined ourselves to the 
linear response limit, wherein the solution is {\em unique} and its lamentable failure
can be unambiguously stated.  
It \emph{is} possible that the DG current vanishes 
beyond the linear response limit,\cite{Baldea:unpublished} but we cannot \emph{safely}
state this. But \emph{if} this were the case, it would have nothing to do with 
the DG claim: as seen above,
constraining populations does not prevent a broken time-reversal symmetry.  
Whether the solution of the nonlinear problem corresponds to $J=0$ or not
did and does not represent our concern once the linear response limit is 
incorrect. This would be yet another unphysical DG prediction, as we already 
stated \emph{explicitly}: see the last but one paragraph 
of Sec.~VI in Ref.~\onlinecite{Baldea:2008b}.

Let us now see what is wrong with the DG argumentation yielding the ``conclusion'' that 
constraining populations is unphysical. There are two reasons why their conclusion is 
incorrect: on one hand, the DG argumentation eliminates an important aspect 
and, on the other hand, it introduces an unjustified assumption. 
DG seem to have realized that current conservation is a serious stumbling block for 
their argumentation;
noteworthy, the steps 1---6 listed in the beginning of the Comment omit the current conservation.
They first discuss the minimization \emph{ignoring} 
the current conservation. 
This amounts to exclude the last term of the square parentheses 
of Eqs.~(\ref{eq-min}), (\ref{eq-An}), and (\ref{eq-An-ast}) above. This \emph{almost} 
suffices to argue that 
Eqs.~(\ref{eq-Q-k}) --- (\ref{eq-An-ast}) automatically imply real coefficients 
$A_n$. ``Almost'' but not \emph{really}, because the unknown $A_n$ can be complex 
($\mbox{Re}A_n \times \mbox{Im}A_n\neq 0$) despite the fact 
that all the other quantities entering these equations are real (remember $x^2+x+1=0$).
Then, to more ``convincingly'' argue that our constraints do not break the time-reversal symmetry,
DG supplement their argumentation with an \emph{unjustified} supposition. Namely,
they simply \emph{postulate} that ``the variational approach assumes a unique solution $\Psi$ \ldots''
The DG variational approach is a well-defined mathematical problem.
Needless to say, whether a solution \emph{exists} and is furthermore \emph{unique} cannot be 
``assumed'' in any mathematical problem. 
This should be \emph{demonstrated}, an imperious requirement
particularly in the case of such an approach, whose shortcomings are so amply documented. 
\cite{Baldea:2008b,Baldea:2009c,Baldea:2010g} In the last part of their argumentation,
DG claim that their deduction that $\Psi$ is real remains
true if they impose current conservation. As discussed above, this unjustified 
assertion 
is contradicted by our Eqs.~(\ref{eq-An}) and (\ref{eq-An-ast}). As a matter of fact,
it is just the opposite sign of the last term in the square 
parentheses of Eqs.~(\ref{eq-An}) and (\ref{eq-An-ast}) that is related to 
the broken time-reversal symmetry expressed by the transformation 
$\Psi \to \Psi^{\ast}$, $J \to -J$ ($I \to -I$ in DG's notation) noted above.
DG also mention this transformation but fail (or avoid) 
to correctly include the current conservation 
as a constraint in the energy minimzation that has to be done.

To summarize, constraining populations is very possible 
and does not in the least exclude a broken time-reversal symmetry.
Were our choice to constrain electronic populations inadequate, which are 
used in the most well-established approaches 
for the uncorrelated and correlated transport 
[like those based on 
Boltzmann's equations, nonequilibrium Green's functions (NEGF), master equations, etc], 
all these approaches would fail, but this is not the case. They all 
employ the Fermi distributions (FDs) 
$f_{\kappa} \equiv \langle \Psi\vert n_{\kappa}\vert \Psi\rangle$
to express open boundary conditions for incoming electrons. 

Before ending this section, we still make the following remark.
Most important for the invalidity of 
the variational DG approach demonstrated in I is that it leads 
(even with the constraints most justifiable physically, namely $n_{\kappa}$) 
to an {\em unphysical} result. 
We paid and pay little attention to the fact that the deduced current vanishes.
By chance, the current could have been nonvanishing, 
as could fortuitously (but must not) 
happen when using the DG variational ansatz with Wigner 
constraints (e.~g., Fig.~3 of Ref.~\onlinecite{Baldea:2008b}).
There are certainly {\em many} possibilities to obtain nonvanishing currents by 
constraining ad hoc complex Hermitian operators, since all these generally yield 
$J \neq 0$: see Eqs.~(18)--(20) of I for $\mathcal{Y}_{\kappa,n} \neq 0$; 
one needs not be too ``careful'' for this, contrary to the impression which DG 
attempt to convey.
Essential to deduce results which are physically relevant is to be 
careful in selecting from the very broad class $\mathcal{C}$ mentioned 
in the Comment those Hermitian operators which are associated to observables
able to express the physical reality at the boundaries. 
And, according to the present community's wisdom, 
a choice better justifiable physically 
than the electronic momentum distributions (FDs) is not known;
DG themselves have to admit that they represent ``an interesting 
alternative to the'' WF (cf.~last paragraph of the Comment).
\section{Refuting the claim that we used real orbitals}
\label{sec-real-orbitals}
Attempting to give a physical basis to their criticism, DG switch to the 
single-particle description and claim that we 
cannot constrain only incoming electrons, 
as appropriate for transport,  
because our single-particle wave functions (orbitals) are real, 
and outgoing electrons become concomitantly constrained.

In I (as well as in Sec.~\ref{sec-time-reversal})
we need not orbitals at all, nor constrain orbitals,
so it is not true what 
DG write that we ``have decided to constrain the occupation of {\em real}
states of the left electrodes of approximate form $\sin(k \pi x/L)$''.
These (approximate) wave functions pertain to an {\em isolated} electrode 
(i.~e., uncoupled to a device) with open boundary conditions (oBCs). 
Here, {\em open} is meant not in the sense of transport theories (of a system 
exchanging particles and energy with environment), but in that used, e.~g., 
in numerical exact diagonalization or DMRG 
(density matrix renormalization group) studies, of a system with open ends 
whose wave function vanishes at the boundaries 
[in DG's notation, $\alpha_{k}^{BK}(x=0) = \alpha_{k}^{BK}(x=L) \equiv 0$]. 
Our boundary conditions (BCs) for the isolated electrodes [cf.~Eq.~(6) of I]
are {\em general} (see Sec.\ \ref{sec-complex}).

Nevertheless, to understand why the DG claim is incorrect
let us suppose in this section that we would have imposed oBCs.
The DG's analysis related to the functions $\alpha_k^{BK}(x)$ explains nothing but 
the trivial fact that a current cannot flow in an isolated electrode.
How could a current flow in an electrode uncoupled to device?
What we constrained is clearly visible in Eq.~(23) of I: 
that the populations of the incoming electrons 
in two \emph{many-body} states ($\vert \Psi_0\rangle$ for $V=0$
and $\vert \Psi\rangle$ for $V\neq 0$) pertaining to the {\em total} system
(electrodes {\em coupled} to the device) are equal.
We did not constrain noninteracting electrons 
confined within an isolated electrode described by the wave function
denoted by DG by $\alpha_{\kappa}^{BK} (x) \approx \sin(2 \pi \kappa/L)$.
This would be meaningless physically.
That our constraints refer to {\em incoming}
electrons was already indicated in the step (ii) of Sec.~II of I.
Our Erratum,\cite{Baldea:2009d} which DG attempt to misinterpret 
to gain more credibility, merely emphasizes
this fact; it neither corrects an error nor retracts a statement. 

In fact, as already noted above and discussed in detail below, our results apply to more general cases than
that analyzed in the Comment, but let us still remain in the latter framework. 
DG calculations would also be possible by using 
single-electron wave functions $\phi_{\kappa}(x; V)$ to build all Slater determinants 
needed to exhaus the total electrode-device Hilbert space 
and express the general 
many-body state $\Psi$ required for the prescribed minimization. (Notice that 
our demonstration of I is exact and does not rely upon any approximate CI 
expansion, contrary to Ref.~\onlinecite{DelaneyGreer:04a}).
The physical content of the orbitals
$\phi_{\kappa}(x, V=0)$ with 
$\phi_{\kappa}(x=q_L, V=0) \equiv 0$ [these are the DG's $\alpha_{\kappa}^{BK}(x)$]
is that incoming ($in$) and outgoing ($out$) electrons 
move symmetrically in the isolated left electrode.
If we formulated in I the DG variational ansatz
in terms of orbitals, the transport approach would have \emph{allowed} 
a broken $in$(coming)-$out$(going)--symmetry when the electrodes are coupled to a device 
[$\phi_{\kappa_{in, out}}(x = q_L; V) \neq 0$] for a nonvanishing bias $V\neq 0$. 
A current $J$ can flow only through a device {\em coupled} to electrodes.
These  $\phi_{\kappa}(x; V)$'s would have been used throughout to express all averages
needed in I [including the rhs of Eq.~(23), that is, at $V=0$ for electrodes \emph{coupled} to a device] 
and \emph{determined} self-consistently from the DG variational ansatz, 
just like our $A_n$'s of I or above in Sec.~\ref{sec-time-reversal}.
Whether this symmetry, \emph{allowed} to be broken, 
\emph{is} indeed broken and a current indeed flows ($J\neq 0$) 
should be determined by performing calculations within the 
DG variational ansatz, and this is exactly what we did within the many-body formalism of I. 
{\em If} the DG variational ansatz 
were valid, the minimization would yield wave functions
$\phi_{\kappa_{in}}(x; V \neq 0)$ 
different from $\phi_{\kappa_{out}}(x; V \neq 0)$ 
and $J \neq 0$, in the same 
way in which it would have led to Im$A_n \neq 0$ in I and in Sec.~\ref{sec-time-reversal} above.
Of course, this cumbersome approach 
based on the first quantization is equivalent to 
the second quantization formalism employed in I, and what is important 
is the unphysical \emph{outcome} ($J=0$). It clearly demonstrates 
that the DG transport ansatz is invalid. 
\section{Rebutting the claim that complex orbitals rescue the DG variational ansatz}
\label{sec-complex}
DG chose the oBCs as a stumbling block for the demonstration of I,
but, because that case turned out to be more subtle, ended becoming 
trapped themselves in this pitfall. By contrast, the case 
of pBCs
indicated by them as a way-out solution is really
much easier to analyze and allows to easily understand the correctness 
of the result $J=0$ of I.

As already noted, our derivation of the unphysical result $J=0$
for the linear response limit in I is very general. 
To give a flavor on how general are the situations where the DG variational ansatz 
fails, we referred to a rather broad class of uncorrelated and correlated models,
which are mostly used in nanotransport, although our demonstration applies well beyond 
that class. 
In I we gave a formal general analytical demonstration and 
needed not bother to do expensive numerical calculations for
certain device models nor become involved by giving a particular type of 
the BCs for the isolated electrodes or explicitly indicating which labels
$\kappa$ refer to single-particle states of incoming electrons and which 
to outgoing electrons, although they can be obviously specified 
(see below in this section and Sec.~\ref{sec-el-h}).

By inspecting the second and third lines of Eq.~(6) of I, 
one can immediately realize that our isolated electrodes can 
be described by \emph{general} boundary conditions. 
Just for not to bother the reader with an unnecessary lengthy 
discussion
of the BCs, we avoided, e.~g., to specify the site indices at the ends 
of the electrodes opposite to the device, and wrote $l \leq q_L$ and $r \geq q_R$ 
in the second and third lines of Eq.~(6) of I. The
message of our Refs.~\onlinecite{Baldea:2008b,Baldea:2009c,Baldea:2010g} is 
unambiguous and leaves no hope: the DG approach is completely incorrect 
and cannot be rescued. Having demonstrated this, we doubted and doubt 
on the usefulness or necessity of a too detailed analysis of 
an approach so incorrect. 
 
The results of I apply for the aforementioned oBCs (Sec.~\ref{sec-real-orbitals}),
but also for boundaries that are, e.~g., periodic (pBCs), anti-periodic (Moebius) 
or general twisted (see, e.~g., Ref.~\onlinecite{Baldea:99b} and citations therein).  
To this, one should keep, e.~g., $M_{R}$ sites in the right electrode and 
impose, e.~g., that the annihilation operators satisfy 
$a_{q_R + M_R} \equiv a_{q_R}$, 
$a_{q_R + M_R} \equiv - a_{q_R}$, or 
$a_{q_R + M_R} \equiv a_{q_R} \exp(i\theta)$
($\theta$ is a real constant phase), respectively.
The left and right electrodes can even be described by 
different types of boundaries. 

In their Comment, DG concede that our idea of 
constraining populations ``is an interesting alternative to the Wigner function'', but 
claim that it is essential to choose complex single-particle wave functions 
associated to pBCs. 
Accepting also this DG's challenge now, let us explicitly work out in detail 
just this case where the pBCs are applied to the isolated electrodes 
and where the orbitals are complex, which DG suggest as a way out to 
rescue their variational ansatz. From now on, we shall always assume 
pBCs unless otherwise specified.
With pBCs, and assuming homogeneous hopping integrals $\mathcal{T}_{L,R}$ in electrodes 
for simplicity, one can write down explicit  
analytical formulas instead of the general unspecified ones of I, 
because the transformation matrix 
$\boldsymbol{\Gamma}$ which diagonalizes $H_{L,R}$ entering Eq.~(22) of I 
is nothing but the well-known Fourier transformation
\begin{eqnarray}
\displaystyle
a_{q_L + l} & = & M_L^{-1/2} \sum_{k}
\alpha_{k} e^{2 \pi i k l/M_L} , \nonumber \\
a_{q_R + r} & = &  M_R^{-1/2} \sum_{p}
\alpha_{p} e^{2 \pi i p r/M_R} .
\label{eq-Fourier}
\end{eqnarray}
Throughout, we use $l = -M_L + 1, \ldots, 1, 0$; $r =0, 1, \ldots, M_R - 1$;
$-M_L/2 \leq k < M_L/2$; $-M_R/2 \leq p < M_R/2$, and assume even $M_{L,R}$ for 
specificity.
With Eq.~(\ref{eq-Fourier}), the Hamiltonians $H_L$, $H_R$, and $H_{D,e}$ 
of Eq.~(6) of I become
($\tau_L \equiv t_{q_L}$, and  $\tau_R \equiv t_{q_R}$)
\begin{eqnarray}
\displaystyle
H_{L} & = & \sum_{k} 
\left[\mu_L - 2 \mathcal{T}_L \cos(2 \pi k/M_L) \right] \alpha_{k}^{\dagger} \alpha_{k} , \nonumber \\
H_{R} & = & \sum_{p} 
\left[\mu_R - 2 \mathcal{T}_R \cos(2 \pi p/M_R) \right] \alpha_{p}^{\dagger} \alpha_{p} , \nonumber \\
H_{D,e} & = & - \tau_L M_L^{-1/2}\sum_{k} 
\left(\alpha_{k}^{\dagger} a_{q_L} + a_{q_L}^{\dagger} \alpha_{k} \right) \nonumber \\
& & - \tau_R M_R^{-1/2}\sum_{p} 
\left(\alpha_{p}^{\dagger} a_{q_R} + a_{q_R}^{\dagger} \alpha_{p} \right) .
\label{eq-H}
\end{eqnarray}
Further operators affected by the transformation (\ref{eq-Fourier}) entering the 
relevant equations of I are
\begin{eqnarray}
\displaystyle
\label{eq-W}
W & = & \frac{eV}{2} \sum_{k} \alpha_{k}^{\dagger}\alpha_{k}
+ \sum_{q} e V_{q} a_{q}^{\dagger} a_{q}
-\frac{e V}{2} \sum_{p} \alpha_{p}^{\dagger}\alpha_{p} , \\
\label{eq-jL}
j_L & = & i \frac{e}{\hbar} \tau_L M_L^{-1/2} \sum_{k} 
\left(\alpha_{k}^{\dagger} a_{q_L} - a_{q_L}^{\dagger} \alpha_{k}\right) , \\
\label{eq-jR}
j_R & = & i \frac{e}{\hbar} \tau_L M_R^{-1/2} \sum_{p} 
\left(\alpha_{p}^{\dagger} a_{q_R} - a_{q_R}^{\dagger} \alpha_{p}\right) ,
\end{eqnarray}
representing the Hamiltonian of the external bias $W$ and the current operators $j_{L,R}$
at the contacts, respectively. 
The sites within the device $q = q_L + 1, \ldots, q_R - 1$ are not affected by 
Eq.~(\ref{eq-Fourier}), and the corresponding operators (device's Hamiltonian $H_D$ 
and the current within the device $j_q$) can be found in I.

By inspecting Eq.~(\ref{eq-H}), one can immediately see that, 
in spite of the fact that the single-particle wave functions 
$\phi_{k}(x_l) \sim \exp(2 \pi i k l/M_L)$ and $\phi_{p}(x_r) \sim \exp(2 \pi i p r/M_R)$
of Eq.~(\ref{eq-Fourier}) are complex and degenerate ($+k$ and $-k$ correspond to the same energy), 
{\em all} the parameters of the Hamiltonian $H = H_L + H_R + H_{D,e} + H_{D}$ 
are real. Consequently, {\em all} its many-body eigenstates $\vert \Psi_{n}\rangle$ 
are (can be chosen) real.
All the parameters entering $W$ are real, so the matrix elements 
$\langle \Psi_n\vert W \vert \Psi_0 \rangle \equiv \mathcal{W}_n$ [Eq.~(13) of 
Ref.~\onlinecite{Baldea:2009c}] are again real. 
As in the general case of I, the matrix elements of the current operator are 
purely imaginary (i.~e, real $\mathcal{J}_{q,n}$). All the matrix elements of the 
particle number operators
$\alpha_{\kappa}^{\dagger} \alpha_{\kappa}$ ($\kappa=p,k$) remain real 
$\langle \Psi_{n}\vert \alpha_{\kappa}^{\dagger} \alpha_{\kappa} \vert \Psi_0 \rangle 
= \mathcal{X}_{\kappa}$ ($\mathcal{Y}_{\kappa} \equiv 0$). 
Eqs.~(24) and (25) of I remain unaltered. Whether imposing current conservation 
(as we did) or not (as DG incorrectly claim that one could do\cite{DelaneyGreer:04a},
as if their method were so good to automatically include current conservation 
\cite{Baldea:2008b,Baldea:2010g}),
the completely unphysical result ($J=0$) follows as the ineluctable conclusion of applying 
the DG variational ansatz. For these pBCs, the labels 
of the single-particle states of incoming \emph{electrons} 
can be explicitly given (see also Sec.~\ref{sec-el-h}): 
$\alpha_{\kappa} \to \alpha_{k_{in}}$ with $0 < k_{in} < M_L/2$
and $\alpha_{\kappa} \to \alpha_{p_{in}}$ with $-M_R/2 < p_{in} < 0$. 
One can now {\em explicitly} see that only these incoming electrons can and are 
to be constrained in Eqs.~(12) and (23) of I or in the present Eq.~(\ref{eq-Q-k}). 
They do differ from the outgoing electrons,
whose labels are $ -M_L/2 < k_{out} < 0$ and $0 < p_{out} < M_R/2$,
and one can convince oneself  {\em explicitly} that outgoing electrons 
(can) remain 
{\em un}constrained. So, these constraints 
correspond to Fig.~7c of Ref.~\onlinecite{Frensley:90} and not to Fig.~7a and 7b,
contrary to what the Comment claims.
By simple algebraic manipulations of Eqs.~(23)--(25) and (15) of I
one can easily \emph{deduce}
that the distributions of the outgoing and incoming electrons are equal, 
$\langle \Psi \vert \alpha_{k}^{\dagger} \alpha_{k} \vert \Psi \rangle 
= \langle \Psi\vert \alpha_{-k}^{\dagger} \alpha_{-k} \vert \Psi \rangle$ 
and 
$\langle \Psi \vert \alpha_{-p}^{\dagger} \alpha_{-p}\vert\Psi\rangle
= \langle\Psi\vert \alpha_{p}^{\dagger} \alpha_{p} \vert \Psi \rangle$.
(Of course, this is not at all surprising in view of the unphysical result $J=0$.)
One can now {\em explicitly} see that the equality of these distributions 
is the {\em outcome} of the calculations of the {\em defective} DG variational ansatz, 
and is 
by no means (not even implicitly) assumed from the very beginning through an inappropriate choice. 
The constraints used by us within the calculations based on the DG variational ansatz
\emph{allowed} a broken symmetry between incoming and outgoing electrons. 
Whether this symmetry is broken or not remained an {\em open} result, which emerged
from the DG transport calculations. 
The \emph{result} is that this symmetry is not broken, $J=0$,
demonstrating the {\em incorrectness} of the DG variational ansatz, and this also 
holds true for the pBCs, contrary to what the Comment argues. From the above analysis of the 
pBCs it is also clear that the electrodes' size $M_{L,R}$ can be arbitrary large (which is 
impossible within the ab initio DG calculations \cite{DelaneyGreer:04a}).
Therefore one can also understand that the unphysical prediction $J=0$ is not limited 
to pBCs but holds for \emph{any} other BCs as well, since otherwise, e.~g., the entire 
philosophy of solid-state physics to apply pBCs would break down.
\section{Wigner function constraints versus particle distribution constraints}
\label{sec-wigner-vs-pop}
DG mention (as we also did) that, by constraining the WF, 
a nonzero current is possible. In Refs.~\onlinecite{Baldea:2008b,Baldea:2009c,Baldea:2010g}, 
we discussed that, \emph{luckily}, mathematically this may be possible. 
Why did DG constrain the WF? Only because in Ref.~\onlinecite{DelaneyGreer:04a} 
they claimed that the FDs 
cannot be used for correlated transport. 
This claim does not at least apply for uncorrelated systems. At least there, 
FD-constraints are possible, and our critique of the DG variational ansatz 
from I obviously applies. 
In fact, in I we explained that open boundary conditions can also 
be formulated by means of 
FDs even for correlated electronic 
devices, because they should be imposed in electrodes, wherein electrons are uncorrelated.
It is the FD which has a precise physical meaning, and not the WF, which has a physical 
content \emph{only} when it is a good approximation for the FD. 
According to the Comment's original philosophy, a current flow is possible 
only to the extent to which the WF does differ from the FD.

The inappropriateness of the Wigner constraints could not be immediately recognized 
only because, luckily, the matrix elements of the Fano operator  
are generally complex. 
They are generally complex no matter whether the single-electron functions 
are real or complex 
(see Refs.~\onlinecite{Baldea:2008b,Baldea:2010g}); this is not the result of 
any ``careful'' choice of certain complex functions, as incorrectly 
claimed by DG. Let us show that this is also the case 
when the Comment's continuous space description is used 
instead of the discrete one of Refs.~\onlinecite{Baldea:2008b,Baldea:2009c}. 
The Fano operator reads
\begin{equation}
\label{eq-fano-general}
F(x, p) \equiv \frac{1}{N}\int d\,r  \ e^{-i p r} \hat{\psi}^{\dagger}(x - r/2) \hat{\psi}(x + r/2) ,
\end{equation}
where $\hat{\psi}^{\dagger}(x)$ and $\hat{\psi}(x)$ the electron field creation and 
destruction operators. 
Its general matrix element for two arbitrary 
$N$-body states $\vert \Psi\rangle$ and $\vert \Phi\rangle$
corresponding to the multielectronic wave functions 
$\Psi(x_1,\ldots,x_N) \equiv \langle x_1 \ldots x_N\vert \Psi\rangle 
\equiv \langle 0 \vert \hat{\psi}(x_1) 
\ldots \hat{\psi}(x_N)\vert \Psi\rangle /\sqrt{N!}$
($\vert 0 \rangle$ is the vacuum)
and 
$\Phi(x_1,\ldots,x_N) \equiv \langle x_1 \ldots x_N\vert \Phi\rangle$
can be easily expressed as ($\mathbf{X}_{N-1} \equiv \{x_1,\ldots,x_{N-1}\}$)
\begin{eqnarray}
 \langle \Phi \vert F(x, p) \vert \Psi\rangle & = & 
\int e^{-i p r}  
\Phi^{\ast}  \left(\mathbf{X}_{N-1}, x - \frac{r}{2}\right)
\nonumber \\
& \times & \Psi\left(X_{N-1}, x + \frac{r}{2}\right) d\,\mathbf{X}_{N-1} d\,r .
\label{eq-fano-matrix}
\end{eqnarray}
This matrix element is generally complex 
irrespective of whether the wave functions $\Psi(\mathbf{X}_N)$ and 
$\Phi(\mathbf{X}_N)$ 
entering Eq.~(\ref{eq-fano-matrix}) 
are real or complex. Eq.~(\ref{eq-fano-matrix}) is general and holds 
whatever the employed single-particle wave functions (which need \emph{not} be specified there).
The complex exponential entering Eq.~(\ref{eq-fano-matrix}) belongs to the 
definition of the Fano operator, Eq.~(\ref{eq-fano-general}),
and has nothing to do with the employed orbitals. Whatever 
the latter, it can be artificially split as
$\exp(-i p r) = \exp[i p (x - r/2)] \times \exp[-i p (x + r/2)]$.
This trivial splitting and the notation $\alpha_k^{DG}(x)$ for these factors 
is obviously done by DG in their Eq.~(7) 
merely for conveying the false impression that complex exponentials would 
represent a key point, which they would have ``carefully'' exploited 
in their work.\cite{DelaneyGreer:04a}

Let us express the population constraints for the incoming electrons 
($p < 0$) of the right electrode ($\mathcal{R}$), Eq.~(23) of I, making use of 
the Fano operator \cite{Baldea:2009c} and Eqs.~(\ref{eq-Fourier})
\begin{equation}
\label{eq-np-fano-pbc}
\sum_{x \in \mathcal{R}} \hspace*{-3.3ex} \int
\langle \Psi \vert F(x,p) \vert \Psi \rangle = \langle \Psi \vert\alpha_{p}^{\dagger} \alpha_{p}\vert \Psi \rangle 
= 
\langle \Psi_0 \vert\alpha_{p}^{\dagger} \alpha_{p}\vert \Psi_0 \rangle =
\sum_{x \in \mathcal{R}} \hspace*{-3.3ex} \int
\langle \Psi_0 \vert F(x,p) \vert \Psi_0 \rangle .
\end{equation}
Instead of constraining the above sums (integrals) 
over $x$ of the Fano operators (populations), DG's discretionary 
constraint of a single term in either electrode (namely, $x=q_{L,R}$)\cite{DelaneyGreer:04a}  
involves a quantity which does not possess a physical meaning.
Emerging from such an \emph{ad hoc mathematical} constraint, 
it is not at all surprising that
the currents predicted by the original DG approach,\cite{DelaneyGreer:04a} 
whether they vanish or not,
are completely unphysical, as demonstrated in Refs.~\onlinecite{Baldea:2008b,Baldea:2010g}.

To conclude, \emph{mathematically} the WF-constraints used in conjunction 
with the DG variational ansatz 
(can) yield a nonvanishing
$J$ (without \emph{any} physical relevance, cf.~Refs.~\onlinecite{Baldea:2008b,Baldea:2010g}) 
even when using real wave functions,
in spite of the claimed ``warning bells''
 that ``real wavefunctions carry no current''.\cite{GreerComment:10}
\section{Constraints for $n$-type and $p$-type conduction}
\label{sec-el-h}
In Sec.~\ref{sec-complex}, to specify the labels of incoming and outgoing electrons, 
we have made the intuitive (or, better, naive) \emph{assumption} that 
the single-particle states with positive (negative) wave numbers $k$ and $p$ 
correspond to right- (left-) motion. However, they are related 
via Eqs.~(\ref{eq-Fourier}) to \emph{quasi}-momenta and not necessarily to 
physical momenta. Especially for later purposes, 
it is important to \emph{demonstrate} that their sign is 
indeed related to the direction of the motion in the \emph{real} world.

Let us consider the isolated left electrode. Its ground state (not to be confounded 
with that of the coupled electrode-device system, 
$\vert \Psi_0\rangle$, cf.~Sec.~\ref{sec-real-orbitals}) is the Fermi sea 
$\vert F\rangle \equiv \left(\prod_{\vert k\vert \leq k_F}
\alpha_{k}^{\dagger}\right)\vert 0 \rangle$, where the Fermi wave vector $k_F$ is 
determined by the number of electrons. Using 
Eq.~(\ref{eq-Fourier}) one can straightfordwardly demonstrate that 
at any position $q_L + l$ within the (left) electrode, the 
average of the electron number current 
$\overline{j}_{q_L + l}^{} = i(a_{q_L + l + 1}^{\dagger} a_{q_L + l} - h.c.)$
vanishes, $\langle F \vert \overline{j}_{q_L + l}^{} \vert F \rangle = 0$.
[Notice the opposite signs of $\overline{j}^{}$ and 
the \emph{electric} current $j$ of  Sec.~\ref{sec-time-reversal} for electrons.]
Let us also consider the states ($M_L/2 > \vert K\vert > k_F$, 
$\vert K^{\prime}\vert \leq k_F < M_L/2$) 
\begin{equation} 
\label{eq-1p-1h}
\left\vert \Phi_{K}^{el}\right\rangle \equiv \alpha_{K}^{\dagger}\vert F\rangle,
\left\vert \Phi_{K^{\prime}}^{h}\right\rangle \equiv \alpha_{K^{\prime}}\vert F\rangle .
\end{equation}
They represent states with one extra 
electron ($el$) and hole ($h$) in the Fermi sea, respectively.
Straightforward calculations using Eqs.~(\ref{eq-1p-1h}) and (\ref{eq-Fourier})
yield
\begin{eqnarray} 
\label{eq-j-1p}
& & \left\langle \Phi_{K}^{el}\left \vert \overline{j}_{q_L + l}^{}\right\vert \Phi_{K}^{el}\right\rangle = + 2\frac{\mathcal{T}_L}{M_L} \sin \frac{2\pi K}{M_L}, \\
\label{eq-j-1h}
& & \left\langle \Phi_{K^{\prime}}^{h}\left \vert \overline{j}_{q_L + l}^{}\right\vert \Phi_{K^{\prime}}^{h}\right\rangle = - 2\frac{\mathcal{T}_L}{M_L} \sin \frac{2\pi K^{\prime}}{M_L} . 
\end{eqnarray}
The sign of the $\overline{j}^{}$-average does \emph{express} 
the real direction of electron quantum-mechanical 
motion. Therefore, Eqs.~(\ref{eq-j-1p}) and (\ref{eq-j-1h})
demonstrate that the sign of the wave vectors belonging to the 
Brillouin zones of Sec.~\ref{sec-complex}
(symmetric around zero) specifies the direction of electron motion,
and that electrons and holes  with a given wave vector move in opposite 
directions. 
The latter result can also be seen by performing the general particle-hole 
transformation, 
$e \to -e$ (charge conjugation) and $\{\hat{\psi}^{\dagger}(x), \hat{\psi}(x)\}$ $\to$
$\{\hat{\psi}_{h}^{\dagger}(x) \equiv \hat{\psi}(x), 
\hat{\psi}_{h}(x) \equiv \hat{\psi}^{\dagger}(x)\}$
($\{a_{x}^{\dagger}, a_{x}\} \to \{a_{x}^{h\, \dagger}  \equiv a_{x}, a_{x}^{h} \equiv a_{x}^{\dagger}\}
$).
Using Eq.~(\ref{eq-fano-general}), one easily gets
\begin{eqnarray} 
\label{eq-j-h}
& & \overline{j}^{h}_{x} = - \overline{j}_x; j^{h}_{x} = + j_x, \\
\label{eq-Fano-h}
& & F_{h}(x, -P) 
\equiv \frac{1}{N}\int d\,r  \ e^{i p r} \hat{\psi}_{h}^{\dagger}(x - r/2) \hat{\psi}_{h}(x + r/2) 
= - F(x, P) + \mbox{const} .
\end{eqnarray}
In view of the aforementioned, one can conclude
that incoming and outgoing \emph{electrons} correspond to the wave vectors
\begin{eqnarray} 
\label{eq-el-in}
& & 0 < k_{in} < M_L/2 ; -M_R/2 < p_{in} < 0 , \\[-2ex]
& &  \mbox{\hspace*{37ex} (for electrons)} \nonumber \\[-2ex]
\label{eq-el-out}
& & -M_L/2 < k_{out} < 0 ; 0 < p_{out} < M_R/2 ,
\end{eqnarray}
while for incoming and outgoing \emph{holes} 
\begin{eqnarray} 
\label{eq-h-in}
& & -M_L/2 < k_{in} < 0 ; 0 < p_{in} < M_R/2 , \\[-2ex]
& &  \mbox{\hspace*{37ex} (for holes)} \nonumber \\[-2ex]
\label{eq-h-out}
& & 0 < k_{out} < M_L/2 ; -M_R/2 < p_{out} < 0 .
\end{eqnarray}
The fact that the above electron and hole descriptions are equivalent 
is trivial in general, but not in the context of 
transport approaches, wherein incoming \emph{charge carriers} are 
to be constrained.\cite{Frensley:90}
If the charge carriers are \emph{electrons} ($n$-type conduction), the constraints 
should be imposed to incoming electrons, 
Eq.~(\ref{eq-el-in}). In this case, the intuitive assumption of 
Sec.~\ref{sec-complex} is justified. 
However, if the charge carriers are \emph{holes} ($p$-type conduction),
one should constrain the incoming holes, Eq.~(\ref{eq-h-in});
that is, the labels $\kappa$ in the above Eq.~(\ref{eq-Q-k}) and in Eqs.~(12) and (23) 
of I are those given by Eq.~(\ref{eq-h-in}) and not by Eq.~(\ref{eq-el-in}).

The analysis of this section and of Secs.~\ref{sec-real-orbitals} and \ref{sec-complex}
makes it now clear why we preferred to consider the general case in I and not to enter 
in unnecessary involved details: they are absolutely not necessary to understand 
the unphysical prediction $J=0$ of the DG variational approach, and 
hence its lamentable failure. But because the incorrect DG claims in the Comment 
brought us to enter such details, we can show another shortcoming of the 
original DG approach related to them, which we did not present so far. 

In their work,\cite{DelaneyGreer:04a} DG did not examine at all whether the molecule 
they considered, BDT (benzenedithiolate), exhibits an $n$- or a $p$-type conduction.
Uncritically, they merely constrained $f(q_{L}, P>0)$ and $f(q_{R}, P<0)$. 
Even if their variational ansatz were correct,
and even if these WFs were true distribution functions, 
these constraints would be appropriate only if the charge carriers were electrons
[conduction mediated by LUMO (lowest unoccupied molecular orbital)].
In reality, in BDT the majority charge carriers are holes ($p$-type conduction), 
as clearly demonstrated 
by the recent, accurate experiment of Ref.~\onlinecite{Reed:09}. 
By inspecting now Eqs.~(\ref{eq-Fano-h}) (noting the reversed sign of 
$P$ in the lhs and rhs), (\ref{eq-el-in}), and (\ref{eq-h-out}),
one is amazed to see that what DG constrained in Ref.~\onlinecite{DelaneyGreer:04a}
are in fact the \emph{outgoing} majority carriers, and \emph{not} the incoming ones.
It is certainly too simplistic to describe the conduction through 
BDT merely as a process mediated by HOMO (highest occupied molecular orbital; $p$-type conduction) 
instead of accounting for several/numerous ionization and electroaffinity levels,
but the fact that the constraints of majority carriers (holes) 
are unphysical in Ref.~\onlinecite{DelaneyGreer:04a} is a clear demonstration that 
uncritically using  
Wigner boundaries is completely unjustified.  
To conclude, even if all the other DG ingredients were correct 
(what is obviously not the case \cite{Baldea:2008b,Baldea:2010g}), this very reason 
irrefutably demonstrates that the results of 
Refs.~\onlinecite{DelaneyGreer:04a,DelaneyGreer:04b,DelaneyGreer:06} 
have absolutely no physical meaning. What would be the appropriate constraints in the case 
of ambipolar conduction,
where both electrons and holes contribute to the current, 
is an issue,\cite{Baldea:unpublished} 
which we do not discuss here.

In our first work \cite{Baldea:2008b} that challenged the DG approach 
with Wigner constraints,\cite{DelaneyGreer:04a}
we considered uncorrelated and correlated quantum dots modeled by
a single level whose energy offset from the electrodes' Fermi level $\varepsilon_F = \mu_L = \mu_R$ 
is $\varepsilon_g$.
The results of the DG calculations presented there (e.~g., in Figs.\ 2--5 and 7) 
are for $\varepsilon_g \geq 0$, that is, the dot's level plays the role of a LUMO 
($n$-type conduction). The charge carriers are electrons, and our constraints
[corresponding to the above Eq.~(\ref{eq-el-in})] refer to incoming electrons.

Both the uncorrelated and the correlated models of Ref.~\onlinecite{Baldea:2008b}
are described by Hamiltonians $H(\varepsilon_g)$ possessing 
a particle-hole (or charge conjugation) symmetry (see, e.~g., 
Ref.~\onlinecite{Baldea:2001a} and citations therein) around $\varepsilon_g = 0$: 
$H_h(\varepsilon_g) = H(-\varepsilon_g)$.
That is, the zero-bias conductance $g(+\varepsilon_g) = g(-\varepsilon_g)$ 
(as well as other relevant properties not considered in 
Ref.~\onlinecite{Baldea:2008b}, e.~g., the whole current-voltage characteristics) 
should be identical irrespective whether the 
level is located above ($+\varepsilon_g$) or 
below ($-\varepsilon_g$) the electrodes' Fermi level $\varepsilon_F$.
Noteworthy, the charge carriers are electrons for positive 
$\varepsilon_g$
and holes for negative $\varepsilon_g$. 
As a test for numerical calculations, 
we checked that DG calculations for the LUMO case ($\varepsilon_g > 0$) constraining
the incoming electrons [Eq.~(\ref{eq-el-in})] and for the HOMO case ($\varepsilon_g < 0$) 
constraining the incoming holes [Eq.~(\ref{eq-h-in})] yield the same, albeit completely 
unphysical linear conductance. 
[The electric current operator has the same sign both in the electron and the hole representation,
cf.~Eq.~(\ref{eq-j-h}).]
As clearly demonstrated,\cite{Baldea:2008b} 
the DG-conductance computed in this way is completely unphysical, but \ldots 
it is still positive, $g_{DG}(\varepsilon_g) \geq 0$ 
both for positive and negative $\varepsilon_g$. 
That is, this (modified) DG approach can
still ``predict'' that electrons flow from the lower potential to the higher potential,
and holes flow from the higher potential to the lower potential.

If we drew the curves of Figs.\ 3, 5, and 7 of Ref.~\onlinecite{Baldea:2008b} 
also for $\varepsilon_g < 0$,\cite{Baldea:unpublished} by blindly
computing the DG conductance using \emph{exactly} 
the DG prescribed constraints,\cite{DelaneyGreer:04a} [i.~e., Eq.~(\ref{eq-el-in})]
we could have shown a funny ``prediction'' of the DG approach \cite{DelaneyGreer:04a}, 
namely, that the linear conductance can be negative, $g_{DG}(\varepsilon_g < 0) < 0$! 
That is, holes should have to flow \ldots 
from the lower potential to the higher potential. This results from the fact 
that the blind constraints of $f(q_L, P>0)$ and $f(q_R, P<0)$ erroneously 
constrain in fact the outgoing carriers; this situation corresponds to Fig.~7d, and not to
Fig.~7c of Ref.~\onlinecite{Frensley:90}. Indeed, these DG Wigner 
constraints break the time-reversal symmetry and yield a nonvanishing current, but 
\ldots what is the physical relevance? 
As a matter of fact, it is just such an unphysical imbalance, which is shown in Fig.~1 (bottom) 
of Ref.~\onlinecite{DelaneyGreer:04a}
(the counterpart of Fig.~7d of Ref.~\onlinecite{Frensley:90} and 
not of Fig.7c, as incorrectly claimed in the Comment), 
that breaks the time-reversal symmetry in Ref.~\onlinecite{DelaneyGreer:04a}.
In Ref.~\onlinecite{Baldea:2008b}, we did not show this conductance $g(\varepsilon_g < 0) < 0$
because the demonstration of the severe failure of the DG approach with Wigner constraints 
was sufficiently convincing even without mentioning this ``prediction'', and 
we preferred to couch the discussion in terms as sober as possible.
However, we have noted it above, since, in spite of the clear evidence 
of Refs.~\onlinecite{Baldea:2008b,Baldea:2010g}, DG 
still continue to uncritically refer to their work \cite{DelaneyGreer:04a} in the Comment.
\section{Further Errors and Inaccuracies in the Comment}
\label{sec-errors}
In the Comment, DG claim that we ``doubt the validity of using the 
Wigner function constraints to apply open system boundary conditions''.
As we repeatedly emphasized,\cite{Baldea:2008b,Baldea:2009c,Baldea:2010g} we did not challenge 
in any of our works published so far \cite{Baldea:2008b,Baldea:2009c,Baldea:2010g}
the imposition of Wigner constraints \emph{in general};
we irrefutably \emph{demonstrated} that applying Wigner constraints 
in the \emph{specific context of the DG variational approach} yields
completely incorrect results. 

Another inaccurate assertion of DG is that ``B\^aldea and K\"oppel 
accept that our {\sl [i.~e., DG's]} Wigner constraints \ldots have \ldots non-zero 
linear response current.'' In reality, we demonstrated 
(i) that even if, \emph{by chance}, 
this is possible,\cite{Baldea:2008b} the DG current 
is completely unphysical, and (ii)
that the DG conductance vanishes \emph{just} in on-resonance cases,
where the exact conductance attains its maximum (unitary limit).\cite{Baldea:2008b,Baldea:2010g}

To refute the false impression, which DG attempt to convey by using the superscript $DG$,
we emphasize that they did never impose pBCs  
for the $Au_{13}$-clusters, which mimic their electrodes
of Refs.~\onlinecite{DelaneyGreer:04a,DelaneyGreer:04b,DelaneyGreer:06}.
For this, it suffices to inspect Eqs.~(1) and (2) of Ref.~\onlinecite{DelaneyGreer:04a}.
Still, let us emphasize for completeness 
that, even if they imposed such Wigner pBCs, the results of 
the DG approach would have been incorrect: in Sec.~V of 
Ref.~\onlinecite{Baldea:2010g}, we demonstrated that the DG conductance 
computed with Wigner constraints vanishes ($g_{DG} = 0$) 
also with pBCs for electrodes, and this occurs just in a typical, 
well-known physical situation 
corresponding to the unitary limit, wherein   
the true conductance reaches the maximum value ($g_{exact} = e^2/h$). 

Our Ref.~\onlinecite{Baldea:2010g} is the only publication wherein the DG approach, 
as originally proposed (i.~e., with WF-constraints), was worked out using plane wave
orbitals $\exp(i\kappa x)$ in  electrodes with pBCs.
DG used nowhere the plane waves $\exp(i\kappa x)$ mentioned in the Comment 
or other orbitals pertaining to electrodes with pBCs, which they
denote by $\alpha_{\kappa}^{DG}(x)$ in the Comment. Or, more precisely,
they should have denoted them so [i.~e., 
$\alpha_{\kappa}^{DG}(x) \equiv \langle x \vert \alpha_\kappa^{DG}\rangle = \exp(i\kappa x)$], 
because in fact their expressions, like $\vert \alpha_\kappa^{DG}\rangle = exp(i \kappa x)$
in the last but one paragraph of the Comment, 
do not comply with textbooks' quantum mechanics.
Concerning the operators denoted by $Q_\kappa^{DG}$, they are useless: 
these operators are not defined at all.

Most significantly, 
the Comment contains equations and assertions, which defy more than eight decades of 
using the second quantization formalism. E.~g.,
DG incorrectly state that the action of the operator $Q_{k}^{BK} = \alpha_{k}^{\dagger BK} \alpha_{k}^{BK}$ 
on a one-electron wave function $\psi(x)$ is ``straight-forward''. 
This operator cannot act at all on $\psi(x)$, not even on a many-electron 
wave function $\Psi(x_1, x_2, \ldots, x_N)$ in the coordinate space, 
as expressed in their incorrect Eqs.~(1) and (2), respectively.
It acts on many-body states $\vert \Psi \rangle$ belonging to the 
abstract Fock vector space.
Further, it is incorrect what they write that the operators $\alpha_{k}^{\dagger BK}$ and 
$\alpha_{k}^{BK}$ create and destroy the {\em eigen}states 
of $H_L$. 
Likewise, these operators act on {\em any} many-body Fock states 
$\vert \Psi_{general}\rangle$, wherein 
they create and destroy one electron in the single-particle state $k$. 
We thank DG for using 
the superscript $BK$ in their notation $\alpha_{k}^{\dagger BK}$ and 
$\alpha_{k}^{BK}$, but it is superfluous. 
The creation and annihilation 
operators $\alpha^{\dagger}$ and $\alpha$ 
used in I [the same as those entering the present Eqs.~(\ref{eq-Fourier})] 
are already unequivocally defined in textbooks, and it is in this sense that 
we used them throughout. 

In view of the aforementioned, the need to translate in the Comment 
obvious results expressed in
the second quantization in I into the first quantization language 
[e.~g., 
the attempt to ``demonstrate'' that the matrix element 
$\langle \Psi_n\vert Q_{\kappa}^{BK}\vert \Psi_0 \rangle$ 
of Eq.~(4) in the Comment is real] is understandable. 
But then it is not surprising 
that understanding what are the actual constraints of I
or the generality of the demonstration of I is not obvious.
Remarkably, although DG 
reproduce our expression of $H_L$ in the Comment, they completely overlook 
the content of this expression: namely, the fact  
the BCs are general, and so are the pertaining orbitals, which are not necessarily real.
Not coincidental is also the fact that DG indicate as a way-out solution just a particular case of I, wherein 
the incorrectness of the DG variational ansatz can be 
immediately understood (cf.~Sec.~\ref{sec-complex}).  

Above, we preferred to respond to and rebut in detail
all the issued raised
by the Comment. In fact, this represents the main objective of our Reply. 
Still, we note that this analysis 
of all the concrete aspects is actually not necessary.
In Sec.~VII of our recent work \cite{Baldea:2010g} we pointed out more serious 
reasons why the DG variational approach fails. 
This criticism also applies to the case discussed above; 
it comprises fundamental aspects not confined to a certain 
type of BCs, let they be in terms of WFs, FDs or others. 
Our criticism was presented in Ref.~\onlinecite{Baldea:2010g} 
in sufficient detail and will not be even summarized in this Reply.
A further fundamental reason why the DG approach is incorrect will be presented 
in Sec.~\ref{sec-isolated}.
\section{The Wigner function is unsuitable to specify the direction of motion}
\label{sec-wigner}
The incorrect claims of DG made us aware of a limitation of the usefulness
of the WF for transport, which we could not find in the literature.

The WF ($f$) is employed in many physical studies, including transport's,
in spite of its physical limitations. 
The limitation known from textbooks \cite{mahan,Datta:97} 
traces back to the Heisenberg's uncertainty principle.
The WF can be negative and should be not interpreted as a probability distribution,
but rather as ``one step in the calculation \ldots never the last step, since''  
is not measurable but ``is used to calculate other quantities that 
can be measured \ldots the 
particle density and current'' and ``no problems are encountered as long as one avoids 
interpreting $f$ as a probability density'' (quotations from 
ch.~3.8, p.~203 of Ref.~\onlinecite{mahan}). 

As noted above, in transport it is helpful
to distinguish between incoming and outgoing electrons. To this aim,
it is necessary to use a physical property enabling to
indubitably assess that electrons are, say, left- or right-moving.
The averages of the 
particle current operator $\overline{j}_x$ used above or the 
{\em physical} momentum do represent such properties. To see whether the
momentum ``variable'' $P$ of the WF $f(x,P)$ justifies to speak of left- or right-moving
electrons depending on the sign of $P$,
let us consider $N$
noninteracting electrons confined within a one-dimensional
square well of width $L$ and infinite height.
(This could be the isolated electrode considered by DG.)
Electrons occupy energy levels $\hbar^2 \kappa^2/(2 m)$, whose
single-electron wave functions
$\phi_{\kappa}(x) = (2/L)^{1/2} \sin(\kappa x) = i (2 L)^{-1/2}[\exp(-i\kappa x ) - \exp(i \kappa x)]$
($\kappa \to \kappa_n = \pi n/L$, $n=1,2,3,\ldots$)
are just those that DG \cite{GreerComment:10} incorrectly claim we would have
used in Ref.~\onlinecite{Baldea:2009c}.
In the ground state, the lowest $N$ levels are occupied up to the Fermi ``momentum''
$p_F = \hbar k_F = \pi \hbar N/L$.
Computing the Wigner function of this system is straightforward:
$f(x, P) = \sum_{\kappa_n \leq \kappa_F}
\int_{x \pm r/2 \leq 0} d\,r \exp(-i P r)\phi_{\kappa_n}^{\ast}(x-r/2)\phi_{\kappa_n}(x+r/2)$
(see, e.~g., Ref.~\onlinecite{mahan}, ch.~3.7, pp.~202-203).
\begin{figure}[h]
\centerline{\includegraphics[width=0.45\textwidth,angle=0]{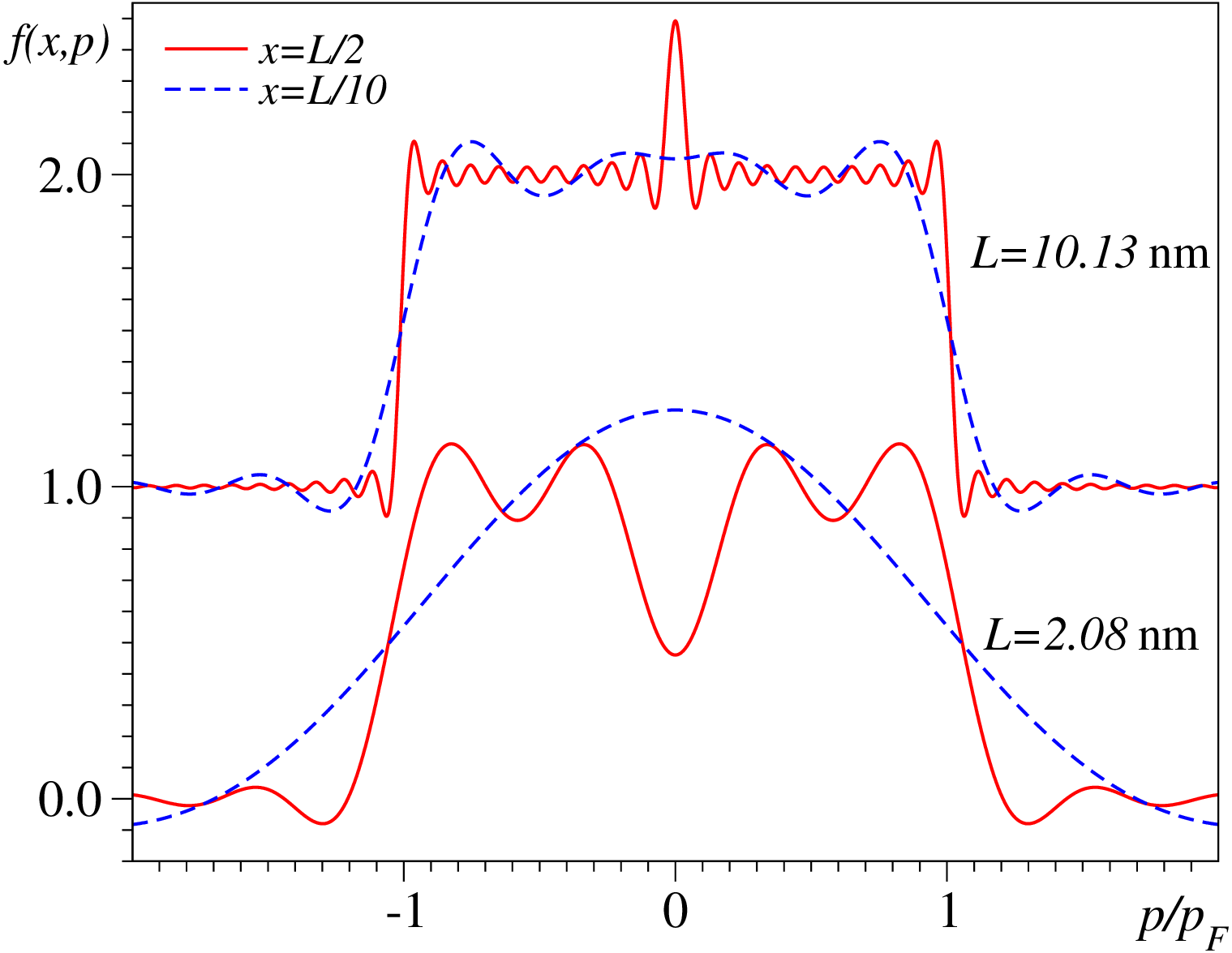}}
\caption{\label{fig:wf} (Color online)
Wigner function for two sizes $L$ computed at two points $x$ indicated in the legend.
Gold's Fermi wave vector $p_F/\hbar=12$\,nm$^{-1}$ is used.}
\end{figure}

One might think that one could use the WF as if it were
a distribution function in cases where its shape resembles a Fermi distribution.
Let us inspect the curves for $f(x, P)$ computed as indicated above and
presented in Fig.~\ref{fig:wf}.
In fact, at smaller sizes
(close to the linear size of the DG's Au$_{13}$-clusters \cite{DelaneyGreer:04a})
the WF does not bear much resemblance to a Fermi function (the lower curves of Fig.~\ref{fig:wf}),
At larger sizes (much larger than those one could hope to tackle within ab initio calculations
to correlated molecules, for which the DG approach \cite{DelaneyGreer:04a} was conceived)
the curves (the upper part of Fig.~\ref{fig:wf}) become more similar to a step function,
and one may think
that this is encouraging. In reality, the contrary is true:
as visible in Fig.~\ref{fig:wf}, \emph{mathematically} one can calculate the WF
for positive and negative ``momentum'' variables $P$ \emph{separately}.
However, this mathematical separation does not reflect a physical reality:
for any single-particle eigenstate $\kappa$
the electron momentum vanishes, $P_{\kappa} \equiv -i \hbar
\int d\,x \phi_{\kappa}^{\ast}(x) \left( \partial/\partial x\right)
\phi_{\kappa}(x) = 0 $;
left- and right-traveling waves are entangled with equal weight, and
one cannot speak of single-particle
eigenstates representing left- or right-moving electrons only because
Wigner functions with positive or negative $P$-arguments can be computed.
This represents a further limitation of the usefulness of the WF,
not related to the Heisenberg's principle,
which is particularly relevant for transport. The current has a direction,
and if one wants to unambiguously specify this direction, the WF $f(x, P)$
is inappropriate;
a WF with negative (positive) ``momentum'', $ P < 0$ ($ P > 0$),
does not imply that left- and right-moving particles exist in the real physical world.

So, using $f(q_L, P>0)$ and $f(q_R, P<0)$ as if they were true 
momentum distributions of 
incoming electrons, as DG did,\cite{DelaneyGreer:04a} is not justified
in quantum mechanics. The above example demonstrates that, indeed, 
the textbook's warning mentioned in the beginning of this 
section is pertinent.
\section{Why any approach to transport merely based on a finite isolated cluster 
necessarily fails}
\label{sec-isolated}
In the course of our extensive critical investigations 
\cite{Baldea:2008b,Baldea:2009c,Baldea:2010g,Baldea:unpublished} 
of the DG approach \cite{DelaneyGreer:04a} we became aware of a series 
of difficulties, which not only the DG's but also other approaches 
to nanotransport are faced with, which we want to bring to 
the reader's attention. 

To understand the problem, let us briefly consider
the uncorrelated dot model of Ref.~\onlinecite{Baldea:2008b} linked to semi-infinite electrodes 
($M_{L,R} \to \infty$).
Exact transport calculations at arbitrary bias $V$ can be easily carried out within 
a multitude of approaches.\cite{Blandin:76,Brako:85,Medvedev:05,Baldea:2010e} 
Besides the current $J$ and dot 
occupancy $N_0$,\cite{Blandin:76,Brako:85,Medvedev:05,Baldea:2010e} the occupancies of the sites 
in electrodes can also be computed.\cite{Baldea:unpublished} 

\begin{figure}[h]
\centerline{\includegraphics[width=0.45\textwidth,angle=0]{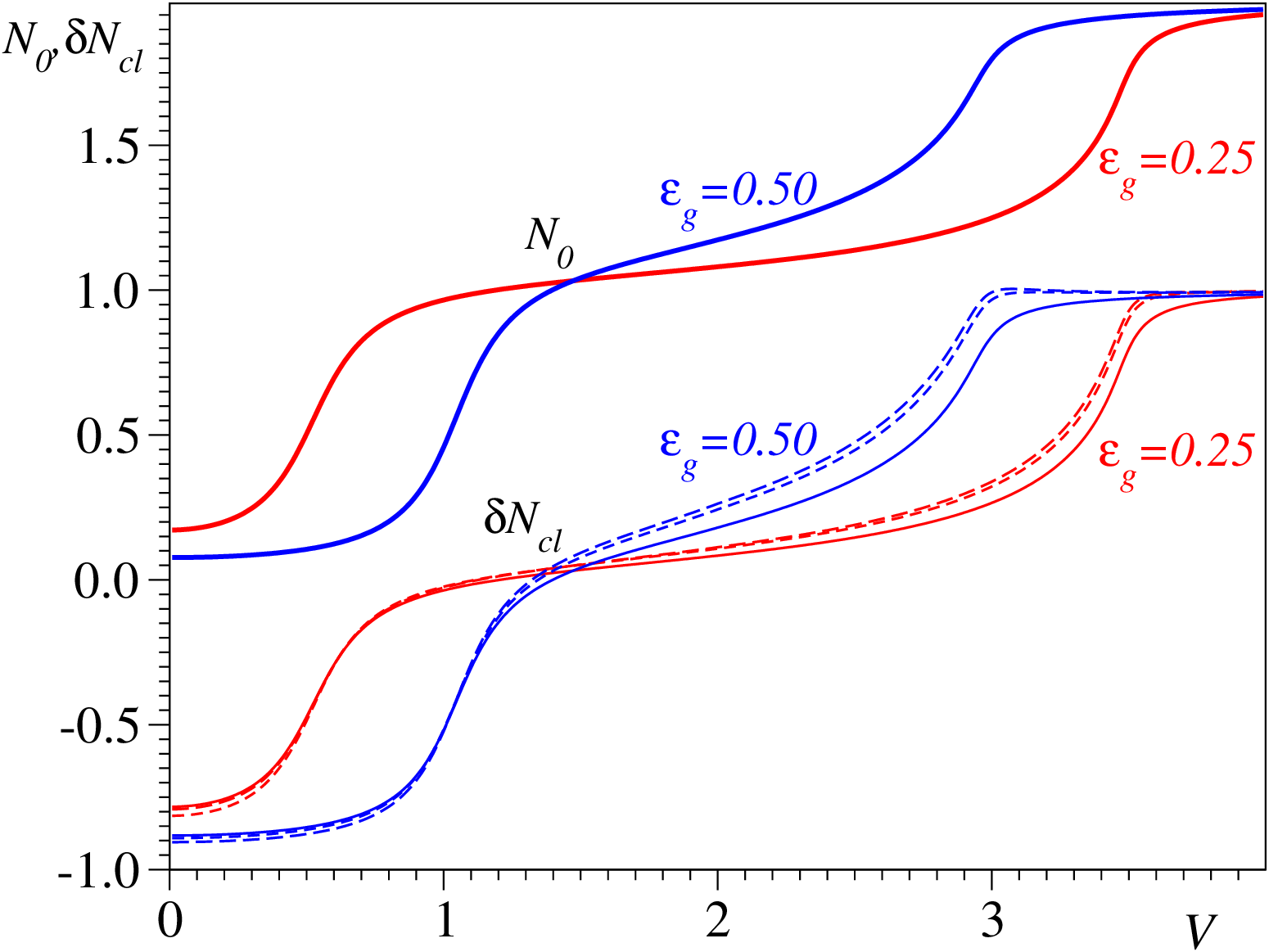}}
\caption{\label{fig:nCluster} (Color online) Dot charge $N_0$ and cluster excess charge 
$\delta N_{cl} = N_{c} - N_s$ for two gate potentials $\varepsilon_g = 0.5$ ($N_s=3,9,11$)
and $\varepsilon_g = 0.25$ ($N_s=3,11,13$). 
For both gate potentials, 
the employed cluster sizes $N_s$ given in parentheses
increase upwards for, say, $V \sim 3$. 
Notice that for larger sizes the excess charge
rapidly saturates and 
closely follows the dot charge, 
and this gives a flavor on how fast the convergence to 
charge neutrality within more distant parts of electrodes is (see the main text). 
Energy unit is $\mathcal{T}_{L,R}=1$ and $\tau_{L,R} = 0.2$.}
\end{figure}

In Fig.~\ref{fig:nCluster}, 
we present steady-state results for the electron number on the dot  $N_{0}$ 
and that for nanoclusters $N_{c}$ centered on the dot and including several 
electrodes' sites $N_s$. As visible there, the 
changes in both $V$ (source-drain voltage) and $\varepsilon_g$ (gate potential) 
yield variations in the dot occupancy $N_{0}$. 
(Notice that there is no exchange of electrons with the gate.)
The total number $N_{c}$ of electrons in the nanoclusters also varies, 
it closely follows the change of $N_{0}$; 
the variations in the small difference $N_{c} - N_{0}$ could 
be hardly seen within the drawing accuracy of Fig.~\ref{fig:nCluster}, 
and therefore are not shown there. The fact that $N_{c} - N_{0}$ is practically 
constant implies that the sites in 
electrodes remain practically unaffected by changes in $V$ and $\varepsilon_g$; 
even the electrodes' sites in the 
very close dot's vicinity are very little affected. 
The dependence $N_{c} \equiv N_{c}(V, \varepsilon_g)$ is most important in the 
context of a transport approach. It demonstrates that, 
by varying $V$ and/or $\varepsilon_g$, a finite nanocluster exchanges electrons with 
the infinite electrodes linked to it, which act as reservoirs that can supply/withdraw electrons. 

In the light of the aforementioned, it becomes clear that it is {\em strictly} 
impossible to correctly describe the transport within a DG-like approach,
which considers a small isolated cluster. 
Within an approach like that initiated by DG \cite{DelaneyGreer:04a} 
and further scrutinized by us 
\cite{Baldea:2008b,Baldea:2009c,Baldea:2010g} 
one attempts to describe the transport using
such a cluster characterized by a many-electron wave function 
$\Psi(x_1, x_2, \ldots , x_N)$ to be determined by 
a constrained minimization of cluster's energy for $V\neq 0$.
This wave function 
merely describes a nanocluster with a given number of electrons $N$, which cannot be 
varied by changing source-drain ($V$) or gate ($\varepsilon_g$) voltages:
\begin{itemize}
\itemsep -0.5ex
\item[(a)] {\em whatever} the property chosen to be constrained 
(WF, electron momentum distribution or else),
\item[(b)] {\em whatever} the boundary conditions (oBC, pBCs, Moebius, general 
twisted or else)
which might be chosen for the finite ``electrodes'' included in the cluster 
(the Au$_{13}$-clusters of Ref.~\onlinecite{DelaneyGreer:04a}), 
\item[(c)]  let 
the many-electron wave function $\Psi(x_1, x_2, \ldots , x_N)$ 
be determined within {\em exact} full CI expansions 
(i.~e, using the whole Hilbert space) as done by us \cite{Baldea:2008b,Baldea:2009c,Baldea:2010g} 
and not only approximately, by using a 
reduced number of configurations selected by Monte-Carlo sampling, as DG did.\cite{DelaneyGreer:04a}
\end{itemize}

Even the most ``carefully'' chosen constraints can at most, e.~g., 
acceptably account for the change the number of electrons in the device (dot).
But this change will inevitably modify the charge of the neighboring sites in electrodes.
It is easy to imagine how profound will be the impact on the transport, e.~g, 
in correlated or switchable devices.
It is hard to conceive that the transport could be reasonably described 
within this framework, even if the cluster would be very large (what 
is obviously hard within the presently available ab initio 
quantum-chemical calculations for correlated systems): because 
it is by no way 
obvious/necessary that the local charge density be incorrectly 
described only at the remote ends of the finite ``electrodes'',
which would allow to suppose/hope that the electric field within the device 
will be little affected.
A further difficulty is, of course, the fact that the number of electrons 
$N_{c}(V, \varepsilon_{g})$ of the coupled nanocluster 
is generally noninteger (cf.~Fig.~\ref{fig:nCluster}), but we do not discuss 
this issue here.\cite{Baldea:unpublished} 

One may further ask whether a ``careful'' choice of a certain charge state of the dot 
(molecule) instead of the neutral species could help. No, it does not. 
For the above model, one can also compute the exact time-dependent dot population $N_{0}(t)$
by suddenly coupling at $t=0$ a dot with population $N_{0}(0)$
to infinite electrodes.\cite{Brako:85,Blandin:76,Medvedev:05}
The result for $V\neq 0$ within the wide band limit ($4\mathcal{T} \gg V,\varepsilon_g, \Gamma$, 
where $\Gamma \equiv 2 \tau^2/\mathcal{T}$, 
$\mathcal{T} = \mathcal{T}_{L} = \mathcal{T}_{R}$, and 
$\tau = \tau_L = \tau_R$) is \cite{Baldea:unpublished}
\begin{eqnarray}
N_{0}(t) & = &
N_{0}(0) e^{- 2 \Gamma t} + 
\left(1 + e^{-2 \Gamma t}\right)  
\label{eq-nD} \\
& & \times \left[1 - \frac{1}{\pi}
\left( \arctan \frac{2\varepsilon_g + e V}{2\Gamma} +
\arctan \frac{2\varepsilon_g - e V}{2\Gamma}
\right)
\right]  \nonumber \\
& & - \frac{2\Gamma}{\pi} e^{-\Gamma t}
\left(
\int_{-\infty}^{-\varepsilon_g - eV/2} + 
\int_{-\infty}^{-\varepsilon_g + eV/2}
\right) d\,\varepsilon \  
\frac{\cos(\varepsilon t)}{\varepsilon^2 + \Gamma^2} . \nonumber
\end{eqnarray}
\begin{figure}[h]
\centerline{\includegraphics[width=0.45\textwidth,angle=0]{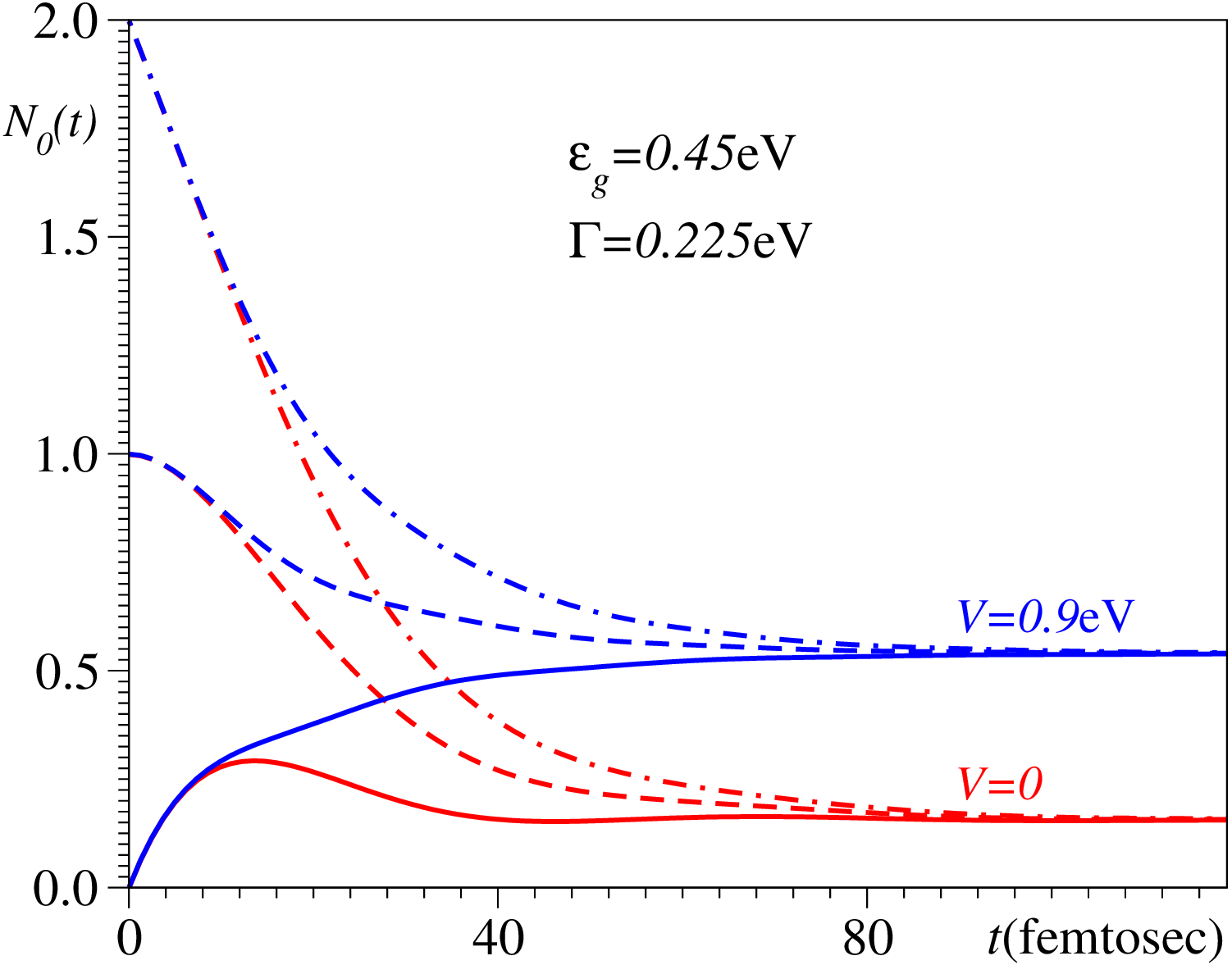}}
\caption{\label{fig:nD} (Color online) Curves for the time-dependent dot charge $N_0(t)$ showing
that the asymptotic values do not keep track of the initial occupancy $N_{0}(0)$. 
The values of the source-drain bias $V$, energy offset $\varepsilon_g$, and hybridisation $\Gamma$
are given in the legend.}
\end{figure}
As seen in Fig.~\ref{fig:nD}, which visualizes the result expressed 
by Eq.~(\ref{eq-nD}), the asymptotic value of the dot population
$\lim_{t\to\infty} N(t)$,
which corresponds to the steady state, does not ``remember'' the initial
value $N(0)$; it is determined by the applied voltages.

The foregoing analysis makes it also clear that not only a
variational ansatz like the DG's is incorrect. The above list (a) -- (c) can be enlarged 
by stating that, {\em whatever} the ansatz (let it be a variational ansatz better than 
the DG's or based on e.~g., Liouville \cite{Frensley:90} or 
Schr\"odinger equations) used to compute the wave function for a fixed $N$, 
it will fail. The aforementioned flaw is serious and can be unambiguously traced back to the use of 
a finite (small) isolated cluster.
Electric transport can only occur in an open system,
and allowing electron exchange with environment (infinite electrodes) 
is indispensable.
Most commonly, this is done via the embedding
self-energies within the Keldysh-NEGF-approach,\cite{Datta:97} 
which enable the nanocluster to change the number of electrons it contains. 
\section{Conclusion}
\label{sec-conclusion}
Most important for the present Reply, 
we have demonstrated above that the DG's claims of the Comment are incorrect and 
our critique of I is untouched.

The unphysical current ($J=0$) obtained in I is the \emph{result} of the DG defective ansatz 
and by no means emerges from pretended unsuitable 
constraints of real orbitals, as DG incorrectly claim.
Using complex orbitals 
does not in the least 
change the unphysical prediction of the DG variational ansatz; furthermore, 
it confirms more directly its incorrectness. In fact, the complex orbitals 
given by DG represents nothing but 
a \emph{particular} case of the general case considered in I, for which our demonstration holds.
For completeness, we have worked out this case in detail to explicitly specify the labels
of the incoming charge carriers (not necessarily electrons), 
to demonstrate that they can be constrained without 
concomitantly constraining the outgoing carriers, 
and to show that the distribution
of the outgoing carrieres represents the \emph{outcome} of transport calculations.
Of course, fully consistent with the unphysical prediction $J=0$, the 
distribution of the outgoing carriers \emph{deduced} within the defective DG ansatz is found 
equal to that of the incoming carriers.

To conclude, DG concede that our idea of 
constraining populations ``is an interesting alternative to the Wigner function'', but 
argue that it is essential for this to choose complex states. 
We have shown above that the usage of the complex orbitals 
\emph{just} of the form indicated by them as key point of a way-out solution
does not mender the DG variational approach. 
These are not only complex, they are {\em just} of the form 
$\alpha_{\kappa}^{DG}(x) = \exp( i \kappa x)$, which 
DG give (and {\em incorrectly} attempt to suggest that they used them
in Ref.~\onlinecite{DelaneyGreer:04a}).  
Consequently, they must now accept that their variational ansatz 
for transport is incorrect.

The analysis done in conjunction 
with the issues raised by DG has led us to reveal two aspects 
of more general relevance for the transport theory.
First, we have presented an example illustrating 
a limitation of the Wigner function $f(x,P)$ 
important for transport, namely that the sign of $P$ is not 
necessarily related to the direction of motion in the real world.
Second, we have presented an important physical reason why  
transport approaches, which, like the DG's,
use information pertaining to a finite isolated cluster are inappropriate.  
This enlarges the basis of our critique recently 
formulated in Sec.~VII of Ref.~\onlinecite{Baldea:2010g}, 
where two other fundamental reasons were exposed.

To summarize our detailed 
investigations on the DG approach,\cite{DelaneyGreer:04a} 
we can state that this approach lamentably fails because virtually 
all its ingredients are incorrect:
\begin{itemize}
\itemsep -0.5ex
\item
The DG approach imposes boundary conditions as if the WF were a true 
particle distribution,
which is not justified quantum mechanically. Replacing the WF-constraints by the 
boundary conditions most justified physically (namely, the Fermi distributions)
attempted in I does not remedy this approach.
\item
Even if the WF were a true momentum distribution, by uncritically constraining 
$f(q_L, P>0)$ and $f(q_R, P<0)$ it is very possible to erroneously 
constrain the \emph{outgoing}
majority charge carriers and not \emph{incoming} ones. 
(This is \emph{just} 
the case in Refs.~\onlinecite{DelaneyGreer:04a,DelaneyGreer:04b,DelaneyGreer:06}.)
\item
The DG approach attempts to describe the transport by using a finite 
cluster
within calculations based on a variational principle 
(entropy maximization), which turned out to be problematic 
even if, unlike in the DG case, the limits of infinite time and infinite 
volume are taken in the correct order.\cite{Bokes:03,Baldea:2010g}
\item
The DG approach aims at describing the transport by means of a wave function
determined for a (small) isolated cluster, whose number of electrons is fixed,
while a nanocluster in a real electric circuit 
does exchange electrons with the infinite electrodes to which it is connected.
\end{itemize}

The lamentable failure of the DG approach was demonstrated 
by explicit calculations for the simplest uncorrelated and correlated, discrete and 
continuous models.\cite{Baldea:2008b,Baldea:2010g} They 
contradict well-established experimental and theoretical results, and it would make 
little sense to more amply document the incorrectness in \emph{many} other cases.\cite{Baldea:unpublished}
By contrast, DG could not present even a single example where it is valid.  

Based on ingredients unfounded physically, it is not at all surprising that
the currents predicted by the DG approach 
are completely unphysical and much poorly agreeing with 
experiment that more common approaches, contrary to the 
seemingly original success claimed in Ref.~\onlinecite{DelaneyGreer:04a}.
In Sec.~VIII of Ref.~\onlinecite{Baldea:2010g}, we clearly showed that 
a standard NEGF-DFT calculation yields currents slightly larger by a factor $\sim 1.5 - 3$, while 
DG's currents \cite{DelaneyGreer:04a} represent $\sim 2 - 5$\% of the experimental currents
of the recent accurate experiment of Ref.~\onlinecite{Reed:09}. 
\section*{Acknowledgment}
The financial support provided by the Deu\-tsche For\-schungs\-ge\-mein\-schaft 
is gratefully acknowledged.
\end{document}